\documentclass[aps,prb,twocolumn,showpacs,showkeys,preprintnumbers,groupedaddress,amsmath,amssymb,]{revtex4-1}

\usepackage{graphicx}
\usepackage{dcolumn}
\usepackage{amsmath}
\usepackage{color}

\begin{document}


\title{Electron dephasing in homogeneous and inhomogeneous indium tin oxide thin films}

\author{Chih-Yuan Wu$^1$}
\author{Bo-Tsung Lin$^2$}
\author{Yu-Jie Zhang$^3$}
\author{Zhi-Qing Li$^3$}
\email[Electronic address: ] {zhiqingli@tju.edu.cn}
\author{Juhn-Jong Lin$^{2,4,}$}
\email[Electronic address: ]{jjlin@mail.nctu.edu.tw}

\affiliation{
$^1$Department of Physics, Fu Jen Catholic University, Hsinchuang 24205, Taiwan\\
$^2$Institute of Physics, National Chiao Tung University, Hsinchu 30010, Taiwan\\
$^3$Tianjin Key Laboratory of Low Dimensional Materials Physics and Preparing Technology, Department of Physics,
Tianjin University, Tianjin 300072, China\\
$^4$Department of Electrophysics, National Chiao Tung University, Hsinchu 30010, Taiwan
}

\date{\today}

\begin{abstract}

The electron dephasing processes in two-dimensional homogeneous and inhomogeneous indium tin oxide thin films have been investigated in a wide temperature range 0.3--90 K. We found that the small-energy-transfer electron-electron ($e$-$e$) scattering process dominated the dephasing from a few K to several tens K. At higher temperatures, a crossover to the large-energy-transfer $e$-$e$ scattering process was observed. Below about 1--2 K, the dephasing time $\tau_\varphi$ revealed a very weak temperature dependence, which intriguingly scaled approximately with the inverse of the electron diffusion constant $D$, i.e., $\tau_\varphi (T \approx 0.3 \, {\rm K}) \propto 1/D$. Theoretical implications of our results are discussed. The reason why the electron-phonon relaxation rate is negligibly weak in this low-carrier-concentration material is presented.

\end{abstract}


\maketitle

\section{Introduction \label{sec 1}}

Indium tin oxide (Sn-doped In$_2$O$_{3-x}$ or so-called ITO) is an interesting and unique material for both fundamental studies and technological applications. For example, good-quality ITO materials exhibit high visible transparencies and low electrical resistivities, \cite{wu4,wu5,wu7} rendering them being widely used in flat displays, solar cells, and resistive touch panes. On the fundamental side, ITO possesses an incomparable free-electron-like energy bandstructure, \cite{wu6,wu8,wu9} while having carrier concentrations ($n$) $\sim 2$ to 3 orders of magnitude lower than those in typical metals. \cite{wu10,wu11,wu13,wu15} As a result, ITO reveals overall metallic behavior whose resistivity versus temperature ($\rho$--$T$) characteristic can be described by the standard Boltzmann transport equation at not too low temperatures. \cite{wu6} At low temperatures, quantum-interference corrections to the temperature dependence of $\rho$ due to the weak-localization (WL) and electron-electron ($e$-$e$) interaction (EEI) effects are pronounced. \cite{wu26} From studies of the WL magnetoresistance (MR), Ohyama and coworkers have measured the electron dephasing length $L_\varphi = \sqrt{D \tau_\varphi}$ in two-dimensional (2D) thin \cite{wu20} and three-dimensional (3D) thick \cite{Ohyama85} \textit{homogeneous} ITO films, where $D$ is the electron diffusion constant, and $\tau_\varphi$ is the electron dephasing time.

Recently, Efetov and Tschersich, \cite{wu18}  and Beloborodov {\it et al.} \cite{wu19} have extensively studied the WL and EEI effects in granular, i.e., \textit{inhomogeneous} metals in the regime where the dimensionless intergrain tunneling conductance $g_T = G_T/(2e^2/h) \gg 1$, where $G_T$ is the intergrain tunneling conductance, and $e^2/h$ is the quantum conductance. The two groups reached a similar conclusion and argued that the WL effect, which was originally formulated for a  homogeneous system, \cite{wu26,wu22} would be suppressed at temperatures $T > T^\ast = g_T \delta/k_B$, where $\delta$ is the mean energy level spacing in the grain, and $k_B$ is the Boltzmann constant. (More precisely, the WL effect would cross over from a global phenomenon to a local single-grain phenomenon.) At temperatures above the characteristic temperature $T^\ast$, only the EEI effect was predicted to be of importance, which, in particular, would dominate the $T$ dependences of both the longitudinal conductivity ($1/\rho$) and the Hall effect. Empirically, Ohyama {\it et al.} \cite{wu20,Ohyama85} had previously found the WL MR in homogeneous ITO films to persist up to temperatures as high as 90 K. Thus, it would be of interest to test whether the WL effect can be observed in inhomogeneous (granular) ITO ultrathin films at $T > T^\ast$. Moreover, this issue is of timely relevance in the light of the recent interest in the electronic transport properties of nanostructured or nanoporous thin films. \cite{RMP07,Huth-jap10}

In the present paper, we report our 2D WL MR measurements on two series of homogeneous and one series of inhomogeneous ITO thin films between 0.3 and 90 K. We found that the WL effect is pronounced in both types of structures and the electron dephasing is governed by the small- and large-energy-transfer $e$-$e$ scattering processes at low and high measurement temperatures, respectively. Below about 1--2 K, depending on samples, the dephasing time became very weakly $T$ dependent. We explain that the fact that the WL MR can be observed at relatively high $T \sim 100$ K is a direct manifestation of the weak electron-phonon ($e$-ph) relaxation in the low-$n$ ITO material. We would like to mention that the homogeneous and inhomogeneous films can be distinguished by their temperature characteristics of the longitudinal resistivity. In homogeneous films, the low-$T$ resistivity ($\rho$) increases with decreasing $T$ as described by the conventional WL and EEI effects. \cite{wu26} On the contrary, in inhomogeneous films, the low-$T$ conductivity (1/$\rho$) reveals logarithm in $T$ dependence as predicted by the recent theories of granular metals \cite{wu18,wu19} (see further discussion in Sec. III).

This paper is organized as follows. Section II contains our experimental method. Section III includes our experimental results and theoretical analysis on the WL MR and the electron dephasing ($e$-$e$ scattering, ``saturated" electron scattering, and $e$-ph scattering) processes. Our conclusion is given in Sec. IV.

\begin{table*}
\caption{\label{Table Wu1} Sample parameters for two series of homogeneous ITO thin films. Films n1 to n4 are as-prepared, while films A1 to A6 are thermally treated. $t$ is the film thickness, $T_a$ is the thermal annealing temperature, ``Gas" is the gas applied during thermal annealing, $R_\square$ is the sheet resistance, $E_F$ is the Fermi energy, $n$ is the carrier concentration, and $D$ is the electron diffusion constant. $1/\tau_\varphi^0$, $A_{ee}^N$, and $A_{ee}$ are defined in Eq.~(\ref{Eq-Fit}). $(A_{ee}^N)^{th}$ and $(A_{ee})^{th}$ are the corresponding theoretical values of $A_{ee}^N$ and $A_{ee}$ calculated according to Eqs.~(\ref{Eq-tau-ee L}) and (\ref{Eq-tau-ee H}), respectively. The thermal annealing time was 1 h. The values of $E_F$ and $n$ were taken from Ref. \onlinecite{wu16}.}

\begin{ruledtabular}
\begin{center}
\begin{tabular}{ccccccccccccc}
Film & $t$ & $T_a$ & Gas & $R_\square$(10\,K) & $E_F$ & $n$(10\,K) & $D$(10\,K) & $1/\tau_\varphi^0$ & $A_{ee}^N$ & $A_{ee}$ & $(A_{ee}^N)^{th}$ & $(A_{ee})^{th}$ \\
& (nm) & ($^{\circ}$C) &  & ($\Omega$) & (eV) & ($10^{20}$\,cm$^{-3}$) & (cm$^2$/s) &($10^9$ s$^{-1}$) & (K$^{-1}$ s$^{-1}$) & (K$^{-2}$ s$^{-1}$) & (K$^{-1}$ s$^{-1}$) & (K$^{-2}$ s$^{-1}$) \\  \hline

n1 & 15 & ---  & ---  & 574 & 0.56 &4.8  &5.7  & 3.1 & 3.3$\times$$10^9$ & $\approx$9$\times$$10^6$  & 9.1$\times$$10^{10}$ & 3.2$\times$$10^7$ \\
n2 & 15 & ---  & ---  & 460 & 0.62 &5.6  &6.8  & 3.0 & 2.8$\times$$10^9$ & $\approx$2$\times$$10^7$  & 7.8$\times$$10^9$    & 2.9$\times$$10^7$\\
n3 & 15 & ---  & ---  & 477 & 0.52 &4.3  &7.1  & 3.3 & 3.0$\times$$10^9$ & $\approx$1$\times$$10^7$  & 8.0$\times$$10^9$    & 3.4$\times$$10^7$ \\
n4 & 21 & ---  & ---  & 100 & 0.74 &7.1 &20 & 11  & 2.8$\times$$10^8$ & $\approx$2$\times$$10^7$  & 2.5$\times$$10^9$    & 2.4$\times$$10^7$ \\
\hline
A1 & 15 & 300  & Ar   & 695 & 0.48 &3.8  &5.1  & 2.2 & 3.4$\times$$10^9$ & $\approx$6$\times$$10^6$  & 1.0$\times$$10^{10}$ & 3.7$\times$$10^7$ \\
A2 & 15 & 500 &  Ar  & 712 & 0.40   & 2.7 &8.2  & 1.6 & 3.6$\times$$10^9$ & $\approx$9$\times$$10^6$  & 1.0$\times$$10^{10}$ & 4.4$\times$$10^7$ \\
A3 & 15 & 550  & Ar   & 813 & 0.37 &2.6  &4.9  & 0.93& 4.1$\times$$10^9$ & $\approx$9$\times$$10^6$  & 1.1$\times$$10^{10}$ & 4.8$\times$$10^7$ \\
A4 & 21 & 400  & O$_2$& 500 & 0.29 &1.9 &6.2  & 1.8 & 3.5$\times$$10^9$ & $\approx$1$\times$$10^7$  & 8.2$\times$$10^9$    & 6.0$\times$$10^7$ \\
A5 & 21 & 450  & O$_2$& 431 & 0.28 &1.6 &7.8  & 2.2 & 3.5$\times$$10^9$ & $\approx$2$\times$$10^7$  & 7.4$\times$$10^9$    & 6.6$\times$$10^7$ \\
A6 & 21 & 450  & Air  & 294 & 0.34 &2.3 &10   & 2.4 & 1.6$\times$$10^9$ & $\approx$1$\times$$10^7$  & 5.6$\times$$10^9$    & 5.1$\times$$10^7$ \\
\end{tabular}
\end{center}
\end{ruledtabular}
\end{table*}

\section{Experiment method \label{sec 2}}

\begin{figure}[htp]
\includegraphics[scale=0.5]{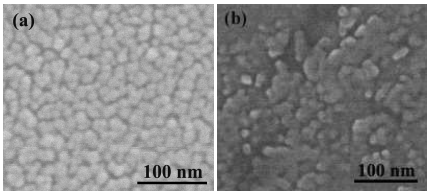}
\caption{(a) An SEM image of the homogeneous n2 film (taken from Ref. \onlinecite{wu17}, see Table~\ref{Table Wu1}), and (b) an SEM image of the inhomogeneous No. 6 film (see Table~\ref{Table Wu2}).} \label{FIG SEM}
\end{figure}

Two series of homogeneous ITO thin films were utilized in this study. The first series of films, with a composition of In$_{1.805}$Sn$_{0.195}$O$_{3-x}$, was 15 nm thick and supplied by the Ritek Corporation. \cite{ritek} The second series of films,  with a composition of In$_{1.835}$Sn$_{0.165}$O$_{3-x}$, was 21 nm thick and supplied by the Aimcore Technology Corporation. \cite{aimcore} All films were fabricated by the rf sputtering method on glass substrates. The crystalline structure of the films was determined by the x-ray diffraction (XRD) measurement, and the film microstructure and morphology were examined by the scanning electron microscopy (SEM). Those studies indicated that the films were composed of relatively small and uniformly distributed grains with average grain size of $\approx 21 \pm 4$ nm [Fig.~\ref{FIG SEM}(a)]. The study also revealed that our films were polycrystalline, rather than amorphous. \cite{wu17} In order to adjust the level of disorder for the WL MR, and thus $\tau_\varphi$, studies, some of the films were thermally treated, as explicitly indicated in Table I. \cite{wu16, wu17} (The samples n1 to n4 stand for a series of as-prepared films, while the samples A1 to A6 stand for a series of thermally annealed films.) In Refs. \onlinecite{wu17} and \onlinecite{wu16}, it had been experimentally demonstrated by some of the authors that these ITO thin films revealed free-carrier-like electronic transport properties. In particular, the electronic parameters, such as the Fermi energy $E_F$ and the elastic mean free path $l$ (mean free time $\tau_e$), have been reliably extracted.

Our inhomogeneous, granular ITO ultrathin films, which had an approximate Sn composition ($\approx$ In$_{1.8}$Sn$_{0.2}$O$_{3-x}$) slightly higher than those of homogeneous films, were also grown by the rf sputtering method on glass substrates. During the deposition process, the mean film thickness $t$, together with the substrate temperature $T_s$, was varied to tune the level of granularity as well as the intergrain tunneling conductivity $g_T$ in each film. The details of the sample fabrication and characterizations had been described in Ref. \onlinecite{wu13} by some of the authors. The SEM studies revealed that our thinnest (5.4 nm) films were porous and possessed notable granular characteristics [Fig.~\ref{FIG SEM}(b)]. On the other hand, there did not exist in the SEM images a clear-cut distinction in the morphologies of our thicker ($\gtrsim$ 9 nm) inhomogeneous films from those of our homogeneous ones. Thus, in this work, we shall adopt the overall $1/\rho$$-$$T$ behavior as a criterion to discriminate the homogeneous from the inhomogeneous ITO films (see Sec. III). Furthermore, it should be stressed that this series of inhomogeneous films, though being granular, still fell deep in the metallic regime, i.e., they had values of $g_T \gg 1$. In Ref. \onlinecite{wu13}, we have experimentally confirmed that the longitudinal conductivity and Hall effect are governed by the EEI effect at temperatures $T > T^\ast$, as recently predicted by Efetov and Tschersich, \cite{wu18}  and Beloborodov and coworkers. \cite{wu19} The relevant parameters of our homogeneous and inhomogeneous samples studied in this work are listed in Tables I and II, respectively. (Remind that our inhomogeneous films are taken from Ref. 9 and the sample names are kept unchanged in order to facilitate possible cross-referencing.)

The low-magnetic-field MR  was measured by using the standard four-probe technique. For the homogeneous films, the samples (typically, $\approx 7$ mm long and $\approx 2$ mm wide) were mounted on the sample holder of an Oxford Instrument Heliox $^3$He cryostat, which was equipped with a 4-T superconducting magnet. The temperature was monitored with calibrated RuO$_2$ and Cernox thermometers. A Linear Research LR-700 resistance bridge with an exciting current of 0.1 $\mu$A was employed to avoid Joule heating. In the case of the inhomogeneous films (typically, $\approx 1$ cm long and $\approx 1.5$ mm wide), the low temperature environment was provided by a physical property measurement system (PPMS, Quantum Design). Both the current source and the voltmeter were provided by the model 6000 PPMS controller. The measuring power limit was set to be 1 $\mu$W. The PPMS was equipped with a 9-T superconducting magnet. In all the MR measurements carried out in this work, the magnetic fields were applied perpendicular to the film plane.

\section{Results and Discussion \label{sec 3}}

Before presenting our WL MR and $\tau_\varphi$ results, we elaborate on the meaning of homogeneous and inhomogeneous films in terms of their low-temperature conductivity versus temperature ($1/\rho$--$T$) behavior. (The SEM studies of the film microstructure and morphology were discussed in Sec. II.) In the case of ``homogeneous" ITO thin films, the low-$T$ resistivity increases with decreasing $T$ and can be well explained within the framework of the conventional WL and EEI effects, as previously shown by Ohyama {\it et al.}, \cite{wu20,Ohyama85} and Lin and coworkers. \cite{wu17,LeeJK} On the contrary, in the case of ``inhomogeneous" ITO ultrathin films, the low-$T$ conductivity obeys a $1/\rho \propto$ ln\,$T$ law over a wide temperature interval from $T^\ast (\approx 2$--3 K) up to several tens K, as recently experimentally established by Zhang and some of the authors. \cite{wu13} It should be emphasized that this logarithmic $1/\rho \propto$ ln\,$T$ law, \cite{wu18,wu19,RMP07} which holds in the presence of granularity and at $T > T^\ast$, is physically distinct from that predicted by the conventional 2D WL and EEI effects. \cite{wu26,wu22} For instance, this newly found $1/\rho \propto$ ln$T$ law in granular metals is independent of sample dimensionality and insensitive to magnetic field. \cite{Sun-prb10} These two kinds of different $1/\rho$--$T$ characteristics apply well to discriminate the homogeneous and inhomogeneous ITO films whose WL MR were studied in this work.

\subsection{Homogeneous ITO thin films}

\begin{figure}[htp]
\includegraphics[scale=0.75]{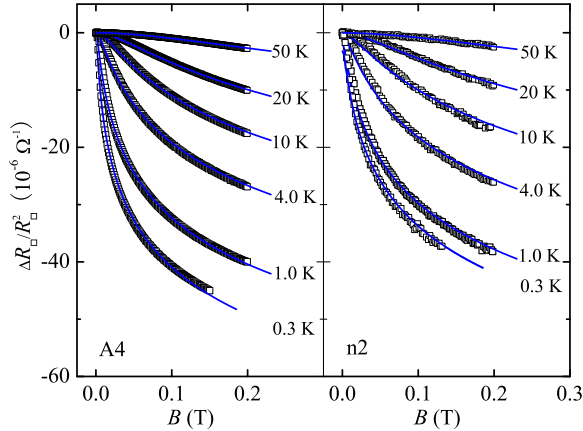}
\caption{(color online) Normalized magnetoresistance $\triangle R_\square  (B)/[R_\square(0)]^2$ as a function of magnetic field at several temperatures for the homogeneous A4 and n2 thin films, as indicated. The magnetic field was applied perpendicular to the film plane. The solid curves are the least-squares fits to Eq.~(\ref{EQ-WL}).} \label{FIG MR-H1}
\end{figure}

We first discuss the WL MR and dephasing time $\tau_\varphi$ in homogeneous thin films. Figure~\ref{FIG MR-H1}  shows the measured normalized MR, $\triangle R_\square (B) /[R_\square (0)]^2 = [R_\square (B) - R_\square (0)] / [R_\square (0)]^2$,  as a function of magnetic field $B$ at several temperatures for the two homogeneous A4 and n2 thin films, as indicated. Here $R_\square = R_\square (T,B)$ is the sheet resistance. This figure indicates that the MR is negative at all temperatures down to 0.3 K, suggesting that the spin-orbit scattering is comparatively weak in ITO. \cite{wu22} (A lower bound of the spin-orbit scattering length/time is estimated below.) On the other hand, the WL MR persists up to above 50 K. The solid curves in Fig.~\ref{FIG MR-H1} are the least-squares fits to the 2D WL theory calculated by Hikami {\it et al.}: \cite{wu21,Altshuler87,Fehr-prb86,Lin-prb87}
\begin{widetext}
\begin{equation}\label{EQ-WL}
\frac{\Delta R_{\square}(B)}{[R_{\square}(0)]^2} = \frac{e^2}{2 \pi^{2} \hbar} \left[ \psi\left(\frac{1}{2} + \frac{B_1}{B}\right)
-\frac{3}{2}\psi\left(\frac{1}{2}+\frac{B_2}{B}\right) +\frac{1}{2}\psi\left(\frac{1}{2}+\frac{B_{\varphi}}{B}\right)
-\ln\left(\frac{B_1}{B}\right) +\frac{3}{2}\ln\left(\frac{B_2}{B}\right) -\frac{1}{2}\ln\left(\frac{B_{\varphi}}{B}\right) \right] \,,
\end{equation}
\end{widetext}
where $\psi$ is the digamma function, and the characteristic fields $B_1 = B_\text{e} + B_{\text{so}} + \frac12 B_0$, $B_2  = B_{\text{i}}+ \frac{4}{3}B_{\text{so}} + \frac{1}{3}B_{0}$, and $B_{\varphi}  = B_{\text{i}}+B_{0}$. The characteristic field $B_j$ is related to the characteristic scattering time $\tau_j$ through the relation $B_j = \hbar/(4eD \tau_j)$, where $j$ = e (elastic scattering time), so (spin-orbit scattering time), i (inelastic scattering time), and 0 (``saturated" dephasing time as $T\rightarrow 0$ K). $2\pi \hbar$ is the Plank constant, $e$ is the electronic charge, and $D$ is the electron diffusion constant. \cite{D-value} Physically, Eq.~(\ref{EQ-WL}) describes how the coherent back-scattering of the electronic time-reversal closed paths in the plane of the film is destroyed by an externally applied small perpendicular magnetic field $B \gtrsim B_\varphi (T)$, leading to the suppression of the WL effect. \cite{wu22} Inspection of Fig.~{\ref{FIG MR-H1}} illustrates that our measured MR can be well described by the predictions of this equation.

\begin{figure}[htp]
\includegraphics[scale=0.75]{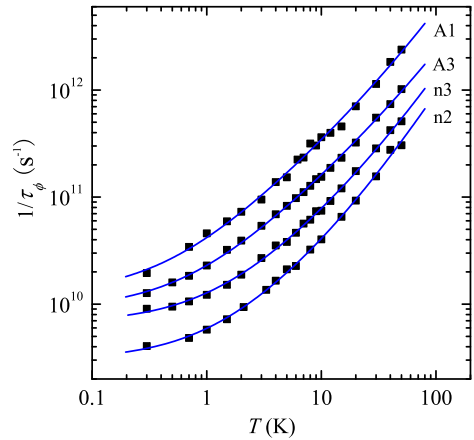}
\caption{(Color online) Electron dephasing rate $1/\tau_\varphi$ as a function of temperature for two as-prepared (n2 and n3) and two thermally treated (A1 and A3) homogeneous ITO thin films. The solid curves are the least-squares fits to Eq.~(\ref{Eq-Fit}). For clarity, the data for the n3, A3, and A1 films have been shifted up by multiplying a factor of 2, 4, and 8, respectively.} \label{FIG Tau-T1}
\end{figure}

Figure~{\ref{FIG Tau-T1}} plots our extracted electron dephasing rate $1/\tau_\varphi$ as a function of temperature for two as-prepared and two thermally treated homogeneous ITO thin films, as indicated. Since our films are 2D with regard to the WL effect (see below for further justification), the responsible dephasing rate is expected to be governed by the $e$-$e$ scattering processes in a significantly wide $T$ interval at low temperatures. Theoretically, the $e$-$e$ scattering rate from the singlet channel has been calculated by Altshuler {\it et al.}, \cite{wu24} and Fukuyama and Abrahams, \cite{wu25} and is given by \cite{wu24, wu25, wu27}
\begin{subequations}
\begin{align}\label{Eq-tau-ee L}
\frac{1}{\tau_{ee}} & = \frac{e^2}{2\pi \hbar^2} R_\square k_BT \ln \left( \frac{\pi\hbar}{e^2 R_\square} \right), & & T<\frac{\hbar}{k_B \tau_e} \\ \label{Eq-tau-ee H}
\frac{1}{\tau_{ee}} & = \frac{\pi}{2} \frac{(k_B T)^2}{\hbar E_F} \ln \left( \frac{E_F}{k_B T}\right). & & T>\frac{\hbar}{k_B \tau_e}
\end{align}
\end{subequations}
Equations~(\ref{Eq-tau-ee L}) and (\ref{Eq-tau-ee H}) stands for the small- and large-energy-transfer $e$-$e$ scattering processes, respectively. Equation~(\ref{Eq-tau-ee L}) is conventionally referred to as the Nyquist quasielastic $e$-$e$ scattering rate, which often dominates in reduced dimensional systems at liquid-helium temperatures. At higher $T$, the second term may be experimentally observed if the $e$-ph relaxation rate is comparatively weak. \cite{Choi-prb87,Eshkol-prb06} (The problem of $e$-ph scattering in a low-$n$ metal is to be discussed in Sec. III E.)

Recently, in addition to the contribution from the singlet channel to the $e$-$e$ dephasing rate, Eqs.~(\ref{Eq-tau-ee L}) and (\ref{Eq-tau-ee H}), the contribution from the triplet channel has been calculated by Narozhny and coworkers. \cite{Narozhny02} In the case of the small-energy-transfer process, Narozhny {\it et al.} found that the triplet channel makes a contribution a factor of $3(F_0^\sigma)^2 /[(1 + F_0^\sigma)(2 + F_0^\sigma)]$ smaller than the singlet-channel term, Eq.~(\ref{Eq-tau-ee L}), where $F_0^\sigma$ is a Fermi-liquid screening parameter. Since the screening parameter is small ($F_0^\sigma \lesssim 0.2$) in the ITO films, \cite{wu17,LeeJK} this contribution is minor and comprises only $\lesssim$ 5\% of the total Nyquist dephasing rate. Similarly, the triplet channel makes a contribution a factor of $3(F_0^\sigma)^2/(1 + F_0^\sigma)^2$ smaller than the rate Eq.~(\ref{Eq-tau-ee H}) to the large-energy-transfer process, which is also a small factor. Therefore, the triplet-channel corrections to $1/\tau_{ee}$ calculated by Narozhny {\it et al.} \cite{Narozhny02} can be safely ignored in the present work.

In Fig.~{\ref{FIG Tau-T1}}, our measured $1/\tau_\varphi$ (the symbols) was least-squares fitted (the solid curves) to the following equation
\begin{equation}\label{Eq-Fit}
\frac{1}{\tau_\varphi (T)} = \frac{1}{\tau_\varphi^0} +A_{ee}^N T + A_{ee} T^2 \ln\left(\frac{E_F}{k_B T}\right),
\end{equation}
where the first, second, and third terms on the right side stand for the ``saturation" term,\cite{Lin-prb87b,Mohanty97,Huang-prl07} the small-energy-transfer term [Eq.~(\ref{Eq-tau-ee L})], and the large-energy-transfer term [Eq.~(\ref{Eq-tau-ee H})], respectively. (The weakly temperature dependent or ``saturated" term $1/\tau_\varphi^0$ will be discussed in Sec. III D.)  Our fitted values of $1/\tau_\varphi^0$, $A_{ee}^N$, and $A_{ee}$ are listed in Table I. This Table also lists the theoretical values of $(A_{ee}^N)^{th}$ and $(A_{ee})^{th}$ calculated according to Eqs.~(\ref{Eq-tau-ee L}) and (\ref{Eq-tau-ee H}), respectively, for each film. Inspection of Table I indicates that our experimental values of $A_{ee}^N$ ($A_{ee}$) are within a factor of $\sim 3$ ($\sim 5$) of the theoretical values. This level of agreement is considered satisfactory. \cite{insensitive}

We note in passing that our extracted electron dephasing length $L_\varphi$ decreases from $\sim 500$ nm at 0.3 K to $\sim 45$ nm at 60 K in all films, justifying that our samples possess strong 2D characteristics with regard to the WL effect. This information, in turn, suggests that the spin-orbit scattering length is $L_{\rm so} = \sqrt{D \tau_{\rm so}} > 500$ nm in our ITO films, corresponding to a spin-orbit scattering time $\tau_{\rm so} > 250$ ps for a film with a diffusivity $D \approx 10$ cm$^2$/s. Similarly, a very weak spin-orbit coupling has recently been found in polycrystalline In$_2$O$_{3-x}$ films \cite{Ovadyahu-prb01}  (where $L_{\rm so}\gtrsim$ 2.8 $\mu$m) as well as polycrystalline SnO$_2$ films. \cite{Dauzhenka-prb11} (In the latter case, the spin-orbit scattering rate was too weak to be experimentally extracted.) The reason why $1/\tau_{\rm so}$ is negligibly weak in these materials, which contain relatively heavy In and Sn atoms, deserves a theoretical explanation.

\begin{figure}[htp]
\includegraphics[scale=0.75]{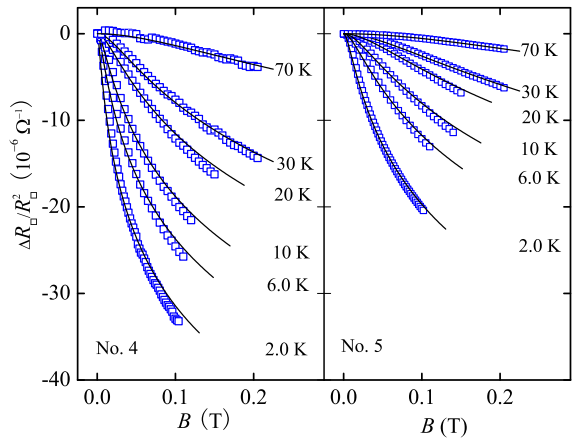}
\caption{(color online) Normalized magnetoresistance $\triangle R_\square (B) /[R_\square (0)]^2$ as a function of magnetic field at several temperatures for the inhomogeneous Nos. 4 and 5 ultrathin films, as indicated. The magnetic field was applied perpendicular to the film plane. The solid curves are the least-squares fits to Eq.~(\ref{EQ-WL}).} \label{FIG MR-H2}
\end{figure}

\begin{table*}
\caption{\label{Table Wu2} Sample parameters for inhomogeneous ITO ultrathin films. $t$ is the mean film thickness, $T_s$ is substrate temperature during deposition, $R_\square$ is the sheet resistance, $E_F$ is the Fermi energy, $n$ is the carrier concentration, and $D$ is the electron diffusion constant. $1/\tau_\varphi^0$, $A_{ee}^N$, and $A_{ee}$ are defined in Eq.~(\ref{Eq-Fit}). $(A_{ee}^N)^{th}$ and $(A_{ee})^{th}$ are the corresponding theoretical values of $A_{ee}^N$ and $A_{ee}$ calculated according to Eqs.~(\ref{Eq-tau-ee L}) and (\ref{Eq-tau-ee H}), respectively. The values of $E_F$ and $n$ were taken from Ref. \onlinecite{wu13}. Note that the $1/\tau_\varphi^0$ values were extrapolated and only approximate.}

\begin{ruledtabular}
\begin{center}
\begin{tabular}{cccccccccccc}
Film & $t$ & $T_s$ & $R_\square$(10\,K) & $E_F$ & $n$(10\,K) & $D$(10\,K) & $1/\tau_\varphi^0$ & $A_{ee}^N$ & $A_{ee}$ & $(A_{ee}^N)^{th}$ & $(A_{ee})^{th}$ \\
 & (nm) & ($^{\circ}$C) & ($\Omega$) & (eV) & ($10^{20}$\,cm$^{-3}$) & (cm$^2$/s) &($10^9$ s$^{-1}$) & (K$^{-1}$ s$^{-1}$) & (K$^{-2}$ s$^{-1}$) & (K$^{-1}$ s$^{-1}$) & (K$^{-2}$ s$^{-1}$) \\  \hline
No.\,1 & 9.7 & 340  &  339 & 0.97 &11 &11 & $\sim$9 & 2.4$\times$$10^9$ & $\approx$1$\times$$10^7$  & 6.2$\times$$10^9$    & 1.8$\times$$10^7$ \\
No.\,2 & 9.2 & 360  &  322 & 0.91 &10 &13 & $\sim$7 & 2.8$\times$$10^9$ & $\approx$1$\times$$10^7$  & 6.0$\times$$10^9$    & 1.9$\times$$10^7$ \\
No.\,3 & 11.3& 380  &  218 & 0.97 &11 &15   & $\sim$7 & 1.8$\times$$10^9$ & $\approx$1$\times$$10^7$  & 4.5$\times$$10^9$    & 1.8$\times$$10^7$ \\
No.\,4 & 13.4& 400  &  160 & 1.03 &12 &17 & $\sim$6 & 1.3$\times$$10^9$ & $\approx$1$\times$$10^7$  & 3.6$\times$$10^9$    & 1.7$\times$$10^7$ \\
No.\,5 & 7.6 & 380  &  687 & 0.74 &7.3 &8.1   & $\sim$8 & 4.0$\times$$10^9$ & $\approx$9$\times$$10^6$  & 1.0$\times$$10^{10}$ & 2.4$\times$$10^7$ \\
No.\,6 & 5.4 & 380  &  1780 & 0.63 &5.8 &4.8 & $\sim$10 & 8.6$\times$$10^9$ & $\approx$8$\times$$10^6$ & 1.8$\times$$10^{10}$ & 2.8$\times$$10^7$ \\
\end{tabular}
\end{center}
\end{ruledtabular}
\end{table*}

\subsection{Inhomogeneous ITO ultrathin films}

As mentioned, recent theories of ``granular metals" (i.e., granular systems with intergrain conductivity $g_T \gg 1$) proposed that the usual WL effect, which was originally formulated for homogeneous systems,  should be suppressed at temperatures $T>T^\ast$ in the presence of granularity. \cite{wu18,wu19} Empirically, it has previously been found that the WL MR in homogeneous ITO films persisted up to several tens degrees of K. \cite{wu20,Ohyama85} In this work, we have measured the low-field MR in a series of granular ITO ultrathin films and compared our results with the WL theoretical predictions, Eq.~(\ref{EQ-WL}).

Figure~\ref{FIG MR-H2} shows our measured normalized MR $\triangle R_\square (B) / [R_\square (0)]^2$ as a function of perpendicular magnetic field for the two Nos. 4 and 5 films at several temperatures, as indicated. It should be noted that the characteristic temperature is known to be $T^\ast = g_T \delta/k_B \approx$ 2--3 K in our granular ITO ultrathin films (see Ref. \onlinecite{wu13}). Figure~\ref{FIG MR-H2} indicates that the measured MR (the symbols) can essentially be described by the 2D WL theoretical predictions of Eq.~(\ref{EQ-WL}) (the solid curves). However, close inspection reveals that Eq.~(\ref{EQ-WL}) is valid only to a narrower magnetic field range ($< 0.1$ T at 2 K, for instant), as compared with that ($\lesssim 0.2$ T at 2 K) in the homogeneous thin films. Hence, the  WL effect, as evidenced in the MR, is not suppressed by the presence of granularity even at $T \gg T^\ast$. This result suggests that the quantum-interference WL effect must cross over fairly smoothly from a global phenomenon to a local single-grain phenomenon as $L_\varphi$ progressively decreases with increasing $T$. This smooth crossover can be ascribed to a relatively long $e$-ph relaxation time in this particular low-$n$ material (see further discussion in Sec. III E). As a consequence of the weak $e$-ph relaxation,  the Nyquist term $A_{ee}^NT$ dominates $1/\tau_\varphi$, and thus $L_\varphi$ varies approximately with $1/\sqrt{T}$ up to comparatively high $T$. Therefore, the WL MR in ITO can still be seen at nearly 100 K, a temperature that is much higher than what could be realized in most normal-metal films. \cite{crossover} (In typical metals, the $e$-ph relaxation usually dominates $1/\tau_\varphi$ already at a few degrees of K, \cite{wu26,wu22,Wu-prb98} causing a much stronger temperature dependent $L_\varphi \propto 1/T^p$ law, with $p \simeq 1$. \cite{wu27,Lin-epl95,Zhong-prl10})

\begin{figure}[htp]
\includegraphics[scale=0.75]{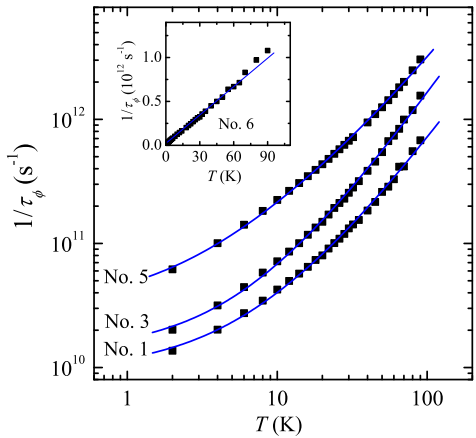}
\caption{(Color online) Electron dephasing rate $1/\tau_\varphi$ as a function of temperature for three inhomogeneous ITO ultrathin films, as indicated. The symbols are the experimental data, and the solid curves are the least-squares fits to Eq.~(\ref{Eq-Fit}). For clarity, the data for the Nos. 3 and 5 films have been shifted up by multiplying a factor of 2 and 4, respectively. Inset: variation of $1/\tau_\varphi$ with $T$ for the No. 6 film plotted in double-linear scales. The straight solid line is a guide to the eye.} \label{FIG Tau-T2}
\end{figure}

Figure~\ref{FIG Tau-T2} shows the extracted $1/\tau_\varphi$ as a function of temperature for three representative inhomogeneous ITO ultrathin films, as indicated. One may see that the magnitudes of $1/\tau_\varphi$ in this case are comparable to those measured in the homogeneous thin films (Fig.~\ref{FIG Tau-T1}). The solid curves in Fig.~\ref{FIG Tau-T2} are the least-squares fits to Eq.~(\ref{Eq-Fit}), and the values of the adjustable parameters $1/\tau_\varphi^0$, $A_{ee}^N$, and $A_{ee}$, together with the corresponding theoretical values  $(A_{ee}^N)^{th}$ and $(A_{ee})^{th}$, are listed in Table~\ref{Table Wu2}. We find that the values of $A_{ee}^N$ and $(A_{ee}^N)^{th}$ agree to within a factor of $\sim 3$ or smaller. Also, the values of $A_{ee}$ and $(A_{ee})^{th}$ are comparable to each other. Thus, both the small- and large-energy-transfer $e$-$e$ scattering processes govern the dephasing in inhomogeneous ITO ultrathin films up to a very high $T \sim 90$ K.

The inset of Fig.~\ref{FIG Tau-T2} indicates that $1/\tau_\varphi$ varies essentially linearly in a wide $T$ range from 2 to $\sim 65$ K in the No. 6 film. This approximate linearity manifests the fact that the small-energy-transfer process [the second term on the right-hand side of Eq.~(\ref{Eq-Fit})] plays a particularly important role in this sample, owing to the high $R_\square$ value ($\simeq 1780$ $\Omega$) of this film.

\subsection{Disorder dependence of small-energy-transfer electron-electron scattering rate}

\begin{figure}[htp]
\includegraphics[scale=0.75]{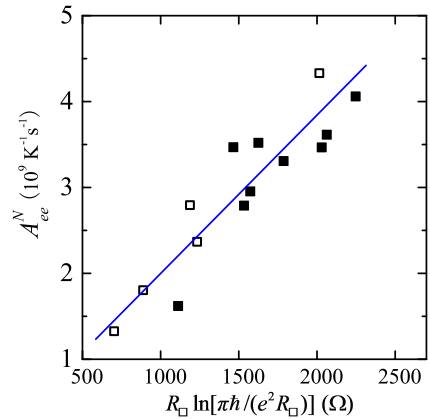}
\caption{(Color online) Extracted small-energy-transfer $e$-$e$ scattering strength $A_{ee}^N$ as a function of $R_{\square}\ln [\pi \hbar /(e^2 R_{\square})]$
for homogeneous (closed symbols) and inhomogeneous (open symbols) ITO thin films. The straight solid line is a guide to the eye.} \label{FIG R_s}
\end{figure}

Figure~\ref{FIG R_s} shows a plot of the variation of our extracted $A_{ee}^N$ values with $R_{\square}\ln [\pi \hbar/(e^2 R_{\square})]$ for both the homogeneous (closed squares) and the inhomogeneous (open squares) ITO thin films. According to Eq.~(\ref{Eq-tau-ee L}), along with the linear $T$ temperature dependence, the small-energy-transfer $e$-$e$ scattering strength $A_{ee}^N$ should vary linearly with the disorder factor $R_{\square}\ln [\pi \hbar/(e^2 R_{\square})]$. This linearity is indeed seen in Fig.~\ref{FIG R_s}, although our data points scatter somewhat. It is worth noting that Fig.~\ref{FIG R_s} includes data points from the 15-nm-thick and 21-nm-thick homogeneous, and the inhomogeneous ITO films, which were obtained from different sources, as described in Sec. II. This result provides a convincing experimental evidence that the Nyquist quasielastic $e$-$e$ scattering process is robust in ITO thin films in a wide $T$ interval from a few K up to several tens degrees of K. This linear law is genuine, especially considering that the value of $R_\square$ of a particular ITO film depends sensitively on the concentration of Sn dopants, the deposition conditions, and (where applied) the thermal treatment conditions.

\subsection{Disorder dependence of ``saturated" electron dephasing rate}

\begin{figure}[htp]
\includegraphics[scale=0.75]{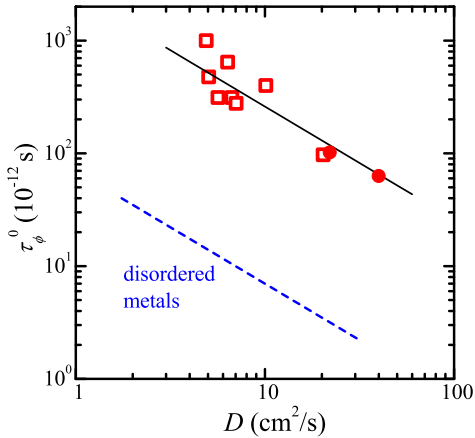}
\caption{(Color online) Variation of $\tau_\varphi^0 = \tau_\varphi (T\rightarrow {\rm 0 \, K})$ with electron diffusion constant for the homogeneous ITO thin films (open squares) listed in Table I. The two closed circles were obtained from two 250-nm-thick ITO films (Ref. \onlinecite{thick}). The straight solid line is drawn proportional to $1/D$ and is a guide  to the eye.  The straight dashed line is taken from Refs. \onlinecite{wu33} and \onlinecite{wu34}, which indicates the experimental $\tau_\varphi^0$ value for typical polycrystalline metals and alloys.} \label{FIG D-T}
\end{figure}

We note that our extracted $1/\tau_\varphi$ values for both homogeneous and inhomogeneous ITO films had to be compared with Eq.~(\ref{Eq-Fit}) which contains a ``constant" term $1/\tau_\varphi^0$. This implies the existence of a very weakly temperature dependent or ``saturated" dephasing time as $T\rightarrow 0$ K. Figure~\ref{FIG D-T} plots the variation of our extracted values of $\tau_\varphi^0$ with the electron diffusion constant $D$ for the homogeneous ITO thin films. \cite{current,tau0}  Surprisingly, this figure illustrates that our extracted $\tau_\varphi^0$ value varies approximately with the inverse of $D$, as is indicated by the straight solid line which is drawn proportional to $1/D$ and is a guide to the eye. Previously, Lin and coworkers \cite{wu33,wu34} have reported an observation of a $\tau_\varphi^0 \propto 1/D^\alpha$ law, with $\alpha$ close to or slightly larger than 1, in numerous highly disordered polycrystalline metals and alloys. This unexpected and seemingly counter-intuitive disorder dependence of $\tau_\varphi^0$ is not yet fully understood. Several distinct theories aiming in addressing this sophisticated problem, while without invoking a magnetic origin, have recently been formulated by, among others, Zawadowski {\it et al.}, \cite{Zawa99} Imry {\it et al.}, \cite{Imry03} Galperin {\it et al.}, \cite{Galperin-prb04} Golubev and Zaikin, \cite{Zaikin07} D\'ora and Gul\'asci, \cite{Dora08} Rotter and coworkers, \cite{wu28} and Chang and Wu. \cite{wu29} In particular, Imry {\it et al.} \cite{Imry03} have proposed that a very weakly temperature dependent dephasing rate in some $T$ interval at low temperatures can arise from specific structural defects in the samples. Indeed, experimental observations of a finite $\tau_\varphi^0$ sensitively associated with metallurgical properties have been reported on a good number of as-prepared \cite{Huang-prl07} as well as thermally treated metal films. \cite{Lin-prb87b,Lin-epl02} Furthermore, Galperin, Kozub and Vinokur \cite{Galperin-prb04} have considered a dephasing model based on tunneling states of dynamical structural defects. They found, under certain conditions, a temperature insensitive dephasing time which obeys an approximate $\tau_\varphi \propto 1/D$ law. In this context, we note that two-level tunneling systems associated with oxygen nonstoichiometries can very likely exist in the ITO material, \cite{Chiu-nano09} and hence be responsible for our observed finite $1/\tau_\varphi^0$.

We would like to note that, for a given $D$ value, the $\tau_\varphi^0$ magnitude for ITO films shown in Fig.~\ref{FIG D-T} is more than 1 order of magnitude longer than the corresponding value (indicated by the straight dashed line) reported in Refs. \onlinecite{wu33} and \onlinecite{wu34} by Lin and coworkers. For example, for a diffusion constant $D \sim 5$ cm$^2$/s, our experimental values are $\tau_\varphi^0 \sim 500$ ps in ITO and $\sim$ 15 ps in polycrystalline metals. Microscopically, it would be of great interest and importance to examine whether a longer $\tau_\varphi^0$ could originate from a reduced Fermi energy $E_F$ (or, equivalently, a low electron concentration \cite{bismuth}) within the theoretical frameworks of Imry {\it et al.}, \cite{Imry03}  and Galperin {\it et al.} \cite{Galperin-prb04} aforementioned. [Check, for instance, Eqs. (4) to (6) in Ref. \onlinecite{Galperin-prb04}.] For comparison, $E_F \lesssim 1$ eV in the ITO material (Tables I and II) while $E_F \gtrsim$ a few eV in typical metals. \cite{Kittel}

Furthermore, we argue that our results shown in Fig.~\ref{FIG D-T} can by no means be ascribed to the magnetic spin-spin scattering, \cite{Micklitz06,Pierre-prb03,Niimi-prb10} because our samples were obtained from different suppliers and thermally treated in different manners, as described in Sec. II. It is hard to conceive that the empirical dependence $\tau_\varphi^0 \propto 1/D$, which has also been found in numerous polycrystalline metals and alloys, \cite{wu27,Huang-prl07,wu33,wu34} as well as in a series of polycrystalline In$_2$O$_{3-x}$ thick films, \cite{Ovadyahu84} could have incidentally arisen from random contamination of dilute paramagnetic impurities. This subtle issue concerning a nonmagnetic dephasing origin calls for further theoretical and experimental investigations. We also would like to point out that Zaikin and coworkers \cite{Zaikin07} have theoretically proposed an $e$-$e$ interaction induced $\tau_\varphi^0 \propto 1/D^\alpha$ law, with $\alpha \approx 1.5$$-$2, for highly disordered metals and alloys having $D$ values smaller than $\sim$ 10 cm$^2$/s.

\subsection{Estimate of electron-phonon relaxation rate in the ITO material}

It is worth examining whether the $e$-ph scattering in ITO might be important and make a contribution to a (nominal) $T^2$ temperature dependence in $1/\tau_\varphi$, which we observed at a few tens K, as described above. An $e$-ph relaxation rate obeying a $T^2$ law is well established for both the dirty limit ($q_Tl < 1$) \cite{Lin-epl95} and the quasi-ballistic limit ($q_Tl > 1$), \cite{Zhong-prl10} where $q_T$ is the wave number of a thermal phonon, and $l$ is the electron mean free path. Theoretically, the electron scattering by transverse vibrations of defects and impurities dominates the $e$-ph relaxation in both regimes. In the $q_Tl >1$ limit (which is pertinent to the present study \cite{qTl}), the relaxation rate, $1/\tau_{e\text{-}t,\text{ph}}$, is given by \cite{Rammer86,wu30,Zhong-prl10}
\begin{equation}\label{Eq-ep}
\frac{1}{\tau_{e\text{-}t,\text{ph}}}=\frac{3\pi^2 k_B^2 \beta_t}{(p_F u_t)(p_F l)}T^2,
\end{equation}
where $\beta_t = (\frac{2}{3}E_F)^2N(E_F)/(2\rho_m u_t^2)$ is a coupling constant, $p_F$ is the Fermi momentum, $u_t$  is the transverse sound velocity, $N(E_F)$ is the electronic density of states  at the Fermi level, and $\rho_m$ is the mass density. Note that Eq.~(\ref{Eq-ep}) predicts a relaxation rate which is proportional to the carrier concentration $n$. \cite{Sergeev} Since ITO has $n$ values which are 2 to 3 orders of magnitude lower than those in typical metals, \cite{wu10,wu11,wu13,wu15} we expect this $1/\tau_{e\text{-}t,\text{ph}}$ rate to be extremely weak.

We may check the possible role of this $e$-ph dephasing term by comparing our measured $1/\tau_\varphi$ (Figs.~\ref{FIG Tau-T1} and \ref{FIG Tau-T2}) with the following equation
\begin{equation}\label{Eq-Fit2}
\frac{1}{\tau_\varphi (T)} = \frac{1}{\tau_\varphi^0} + A_{ee}^N T + \tilde{A}\, T^2 \,,
\end{equation}
where $1/\tau_\varphi^0$ and $A_{ee}^N$ have similar meaning to that in Eq.~(\ref{Eq-Fit}), and $\tilde{A}$ is an adjustable parameter which presumably specifies the $e$-ph scattering strength given in Eq.~(\ref{Eq-ep}). As may be expected, our measured $1/\tau_\varphi$ in each film could be reasonably described by Eq.~(\ref{Eq-Fit2}) (not shown), with an average fitted value of $\tilde{A} \sim 5 \times 10^7$ K$^{-2}$ s$^{-1}$.

The theoretical magnitude of Eq.~(\ref{Eq-ep}) can be fairly reliably evaluated for our ITO films, because ITO possesses a unique free-carrier-like bandstructure and the $E_F$ value in each film (see Tables I and II) has been directly measured through the electronic diffusive thermopower. \cite{wu16} Substituting $u_t \approx 2400$ m/s, $\rho_m \approx 7100$ kg/m$^3$ (Ref. \onlinecite{thin}), and the relevant electronic parameters into Eq.~(\ref{Eq-ep}), we obtain a typical $e$-ph relaxation rate of $\approx 3 \times 10^6 \, T^2$ s$^{-1}$. This relaxation rate is about 20 times smaller than our fitted $\tilde{A}T^2$ value. Therefore, the $e$-ph scattering process, Eq.~(\ref{Eq-ep}), cannot play any significant role in causing the electron dephasing in this work. In this context, previous studies of $1/\tau_\varphi$ in polycrystalline In$_2$O$_{3-x}$ films \cite{Ovadyahu84,Ben-Shlomo-prb89} [where $n \approx$ (5$-$9) $\times 10^{19}$ cm$^{-3}$] revealed no signature of $e$-ph relaxation up to a relatively high $T \approx$ 50 K. In addition, a recent measurement on polycrystalline SnO$_2$ films \cite{Dauzhenka-prb11} (where $n \sim 5 \times 10^{18}$ cm$^{-3}$)  found that the small-energy-transfer $e$-$e$ scattering dominated $1/\tau_\varphi$ up to at least 10 K (no data were reported for higher temperatures). These results are supportive of our assertion in the present work that the $e$-ph relaxation is extremely weak in low-$n$ conductors.

For comparison, we have previously extracted a much higher value of $1/\tau_{e\text{-}t,\text{ph}} \approx 5 \times 10^9 \, T^2$ s$^{-1}$ in the typical alloy AuPd. \cite{Zhong-prl10} In short, the fact that the WL MR can be measured up to nearly 100 K in ITO is a direct manifestation of the relatively weak $e$-ph scattering in this particular low-$n$ material. \cite{ITO-NWs}

\section{Conclusion}

We have extracted the electron dephasing time in homogeneous and inhomogeneous indium tin oxide thin films from the two-dimensional weak-localization magnetoresistance studies. We found that the dephasing rate was governed by the small-energy-transfer electron-electron scattering process at low temperatures, crossing over to the large-energy-transfer electron-electron scattering process at several tens K. The reason why the inelastic electron-phonon relaxation is negligibly weak was explained. At our lowest measurement temperatures, a very weak temperature dependent or ``saturated" dephasing time, which scaled with the inverse of the diffusion constant, was observed. This saturated term may result from electron coupling with specific dynamical structural defects.

\begin{acknowledgments}

The authors are grateful to S. P. Chiu and J. K. Lee for helpful experimental assistance, and I. S. Beloborodov, Y. M. Galperin, A. Sergeev and A. Zawadowski  for valuable discussion. This work was supported by the Taiwan National Science Council through Grant Nos. NSC 99-2120-M-009-001 and NSC 100-2120-M-009-008, and the MOE ATU Program (J.J.L.), and by the NSF of China through Grant No. 11174216, the Key Project of Chinese Ministry of Education through Grant No. 109042, and Tianjin City NSF through Grant No. 10JCYBJC02400 (Z.Q.L.).

\end{acknowledgments}


\begin{thebibliography}{00}
\bibitem{D-value} In this work, our $D$ values were calculated by using an effective electron mass $m^\ast \approx 0.4 \, m_e$ (Ref. \onlinecite{wu6}), where $m_e$ is the free electron mass.
\bibitem{insensitive} The larger discrepancy (a factor of $\sim$ 5) between the $A_{ee}$ and $(A_{ee})^{th}$ values partly arose from the fact that this term contributed only a few percent to the measured $1/\tau_\varphi$. Thus, the least-squares fits were less sensitive to this term. Furthermore, according to Ref. \onlinecite{Narozhny02}, the prefactor $\pi/2$ in Eq.~(\ref{Eq-tau-ee H}) should be replaced by $\pi/4$. This would bring our experiment into a closer agreement with the theory.
\bibitem{crossover} The fact that the WL MR persists up to $T \gg T^\ast$ can be reasoned as follows. In Refs. \onlinecite{wu18} and \onlinecite{wu19}, $T^\ast$ is defined as the temperature where the effective thermal diffusion length in the EEI effect is $L_T^\ast (T^\ast) = \sqrt{D \hbar /k_BT^\ast} \approx a$,  where $a$ is the size of the grain. Since the inequality $L_\varphi > L_T^\ast$ generally holds in (granular) metals, one would have to raise to another higher characteristic temperature, $T_{\rm WL}^\ast$, to reach the condition $L_\varphi (T_{\rm WL}^\ast) = \sqrt{D \tau_\varphi (T_{\rm WL}^\ast)} \approx a$. That is, the WL effect would cross over from a global feature to a local single-grain feature at $T \approx T_{\rm WL}^\ast > T^\ast$. (Here we have assumed that the effective diffusivity for intergrain conduction is $D^\ast \approx D$, because we focus on the regime $g_T \gg 1$.) Numerically, we have $T^\ast \sim 3$ K, and $L_\varphi (T^\ast) \sim 250$ nm $>$ $L_T^\ast (T^\ast) \sim 50$ nm in a film with a diffusivity $D \sim 10$ cm$^2$/s. The approximate equality $L_\varphi (T_{\rm WL}^\ast) \approx L_T^\ast (T^\ast)$ will occur around $T_{\rm WL}^\ast \sim 100$ K.
\bibitem{current} It should be stressed that our excitation current used for the WL MR measurements satisfied the condition $eV_\varepsilon \ll k_BT$, where $V_\varepsilon$ is the applied voltage across the energy relaxation length. See, for example, Z. Ovadyahu, Phys. Rev. B \textbf{63}, 235403 (2001).
\bibitem{tau0} The WL MR in our inhomogeneous ultrathin films were measured only down to 2 K. Thus, we shall not elaborate on the $\tau_\varphi^0$ feature in this case. The physics underlying the $\tau_\varphi^0$ behavior in an inhomogeneous system could be even more complex than that in a homogeneous system.
\bibitem{qTl} For instance, the product $q_T l \approx (k_BT/\hbar u_t)l \approx 0.22 \, T$ for a typical ITO film with $l \approx 4$ nm. This $q_Tl$ value is sufficiently greater than unity at several tens degrees of K.
\bibitem{ITO-NWs} In Ref. \onlinecite{wu15} on the studies of single-crystalline ITO nanowires, the authors had tentatively ascribed their (nominal) $T^2$ term in the measured $1/\tau_\varphi$ to the $e$-ph relaxation rate. Following the analysis of the present work, we learned that the $T^2$ term initially reported in Ref. \onlinecite{wu15} would most likely originate from the large-energy-transfer $e$-$e$ scattering process, Eq.~(\ref{Eq-tau-ee H}).
\end{thebibliography}
\end{document}